\documentclass[journal]{IEEEtran}

\IEEEoverridecommandlockouts

\usepackage{algorithm}
\usepackage{multirow}
\usepackage{booktabs}
\usepackage{cite}
\usepackage{amsmath,amssymb,amsfonts}
\usepackage{algorithmic}

\usepackage{graphicx}
\usepackage[caption=false,font=footnotesize,labelfont=rm,textfont=rm]{subfig}
\usepackage{textcomp}
\usepackage{xcolor}
\usepackage{epstopdf}
\usepackage{cuted}
\usepackage{capt-of}
\usepackage[colorlinks,linkcolor=blue,anchorcolor=blue,citecolor=blue]{hyperref}
\usepackage{orcidlink}

\def\BibTeX{{\rm B\kern-.05em{\sc i\kern-.025em b}\kern-.08em
    T\kern-.1667em\lower.7ex\hbox{E}\kern-.125emX}}

\begin{document}

\title{Enhancing FANET Routing Resilience: A Fuzzy-Driven Bio-Inspired Approach and Its Quantitative Evaluation}

\author{Xinwang~Yuan$^{\orcidlink{0000-0001-8978-876X}}$, 
        Jinshu~Su$^{\orcidlink{0000-0001-9273-616X}}$, 
        Yusheng~Xia$^{\orcidlink{0000-0003-4290-7279}}$, 
        and~Congxi~Song$^{\orcidlink{0000-0002-7672-0915}}$
\thanks{Corresponding author: Jinshu Su.}
\thanks{The authors are with the College of Military Intelligence, Academy of Military Science, Beijing, China (e-mail: yuanxinwang@alumni.nudt.edu.cn; sjs@nudt.edu.cn; xys@nudt.edu.cn; songcongxi17@nudt.edu.cn).}
\thanks{This work is supported by the National Natural Science Foundation of China (Grant No.~62372462).}
\thanks{This work has been submitted to the IEEE for possible publication. Copyright may be transferred without notice, after which this version may no longer be accessible.}}

\markboth{Preprint --- Submitted to IEEE Internet of Things Journal}%
{Yuan \MakeLowercase{\textit{et al.}}: Enhancing FANET Routing Resilience}

\maketitle

\begin{abstract}
In flying ad hoc networks, the high mobility of unmanned aerial vehicles leads to rapid topology changes and unstable links. Clustering can simplify topology management, yet maintaining stable cluster structures remains challenging in highly dynamic environments. This paper employs multi-factor optimization to enhance the artificial bee colony-based cluster head election algorithm and introduces a dual-layer beacon mechanism for cluster maintenance, thereby improving cluster stability and reducing reconfiguration overhead. Meanwhile, although shortening the hello interval helps capture link variations in a timely manner, it significantly increases control overhead. To address this issue, a fuzzy logic-based hello interval controller is proposed to adaptively adjust control message frequency according to network dynamics, effectively reducing overhead while preserving transmission performance. Furthermore, to quantitatively evaluate routing resilience in large-scale and highly dynamic scenarios, this paper proposes a parameter sensitivity index to characterize both overall protocol performance and its sensitivity to environmental variations. The proposed protocol is compared with common baseline schemes in NS-3 simulations, and the resilience evaluation framework analysis reveals significant advantages in overall performance while exposing the trade-off between peak performance and parametric stability.
\end{abstract}

\begin{IEEEkeywords}
Flying ad hoc network, Unmanned aerial vehicle, Routing protocol, Clustering, Fuzzy logic, Performance evaluation
\end{IEEEkeywords}

\section{Introduction}\label{sec:Intro}

\begin{figure*}[htbp]
\centerline{\includegraphics[width=5.8in]{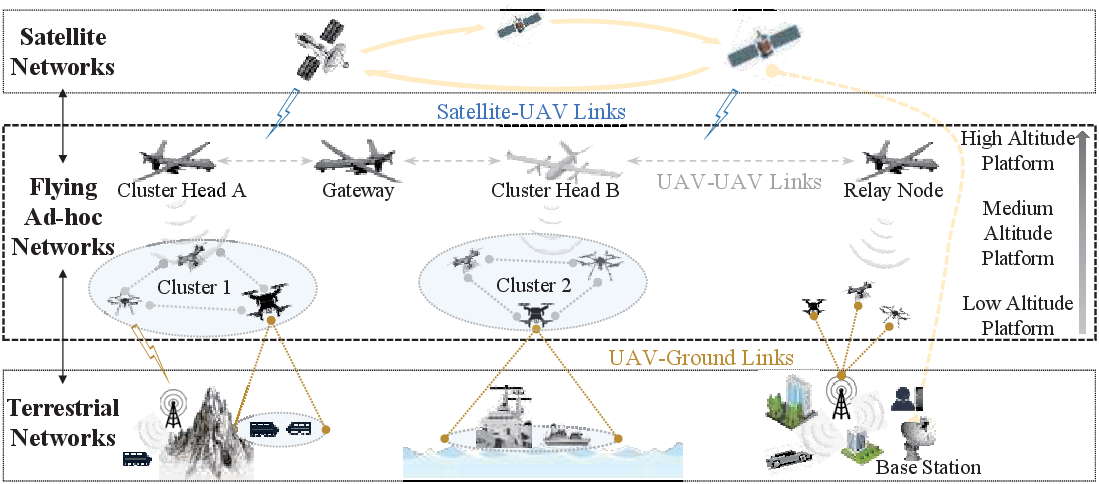}}
\caption{FANET spatial architecture.}
\label{fig:fanet_archit}
\end{figure*}

\IEEEPARstart{F}{lying} ad hoc networks (FANETs) have emerged as a key enabling technology for unmanned aerial vehicle (UAV) communications in highly dynamic scenarios such as precision agriculture~\cite{UAVagriculture}, disaster response~\cite{disaster}, and environmental monitoring~\cite{environmentMonitor}. As illustrated in Fig.~\ref{fig:fanet_archit}, FANETs operate between terrestrial and satellite networks, serving as an intermediate communication layer. Compared with ground-based networks, FANET exhibits significantly higher node mobility, volatile 3D topologies, and more stringent energy constraints. Meanwhile, the ongoing miniaturization and cost reduction of UAV platforms are driving deployments toward increasingly large swarms, further intensifying challenges in scalability, load balancing, and control overhead. As a result, conventional routing protocols like AODV and OLSR degrade in performance due to static control intervals and frequent route reconstructions.

Cluster-based routing has been extensively investigated as a natural approach to managing such large-scale networks by organizing UAVs into hierarchical structures that simplify link management. To enhance cluster-head stability, election mechanisms have evolved from single-factor strategies based on distance~\cite{K-MORP} or energy~\cite{IMRL} to multi-factor evaluations considering neighbor degree, residual energy, and mobility similarity~\cite{icra,pica}. Recent advancements have further shifted towards bio-inspired swarm intelligence~\cite{IABC,MGO_JAYA} and multi-criteria decision-making strategies~\cite{icra,pica,MWCRSF}.
However, existing approaches still struggle with passive cluster maintenance, the absence of adaptive control signaling, and insufficient resilience-oriented evaluation metrics. To address these issues, we propose FBCR, a resilient and scalable clustered routing with adaptive hello beacon controller, along with a novel evaluation framework to validate routing performance and resilience under dynamic scenarios. The technical challenges and our corresponding solutions are detailed as follows:

\textbf{Insufficient static metrics for resilience in dynamic environments.} Existing FANET evaluations primarily rely on conventional QoS metrics. However, these metrics are insufficient to characterize the adaptability and robustness of routing protocols in highly dynamic aerial environments where network scale and node mobility vary significantly. We formulate a resilience evaluation framework to reveal the performance robustness of routing algorithms against environmental parameter variations.

\textbf{Lack of proactive cluster structure maintenance.} While existing studies incorporate basic maintenance like backup cluster heads or merging~\cite{pica,SIC}, their structural management remains passive, relying heavily on unidirectional beaconing or periodic global re-clustering~\cite{IABC,K-MORP,FMORT}. To address this, we design a dual-layer beacon cluster maintenance mechanism to facilitate timely, proactive membership restoration by enforcing continuous bidirectional monitoring, thereby avoiding frequent global re-clustering.

\begin{figure}[!h]
\centerline{\includegraphics[width=9cm]{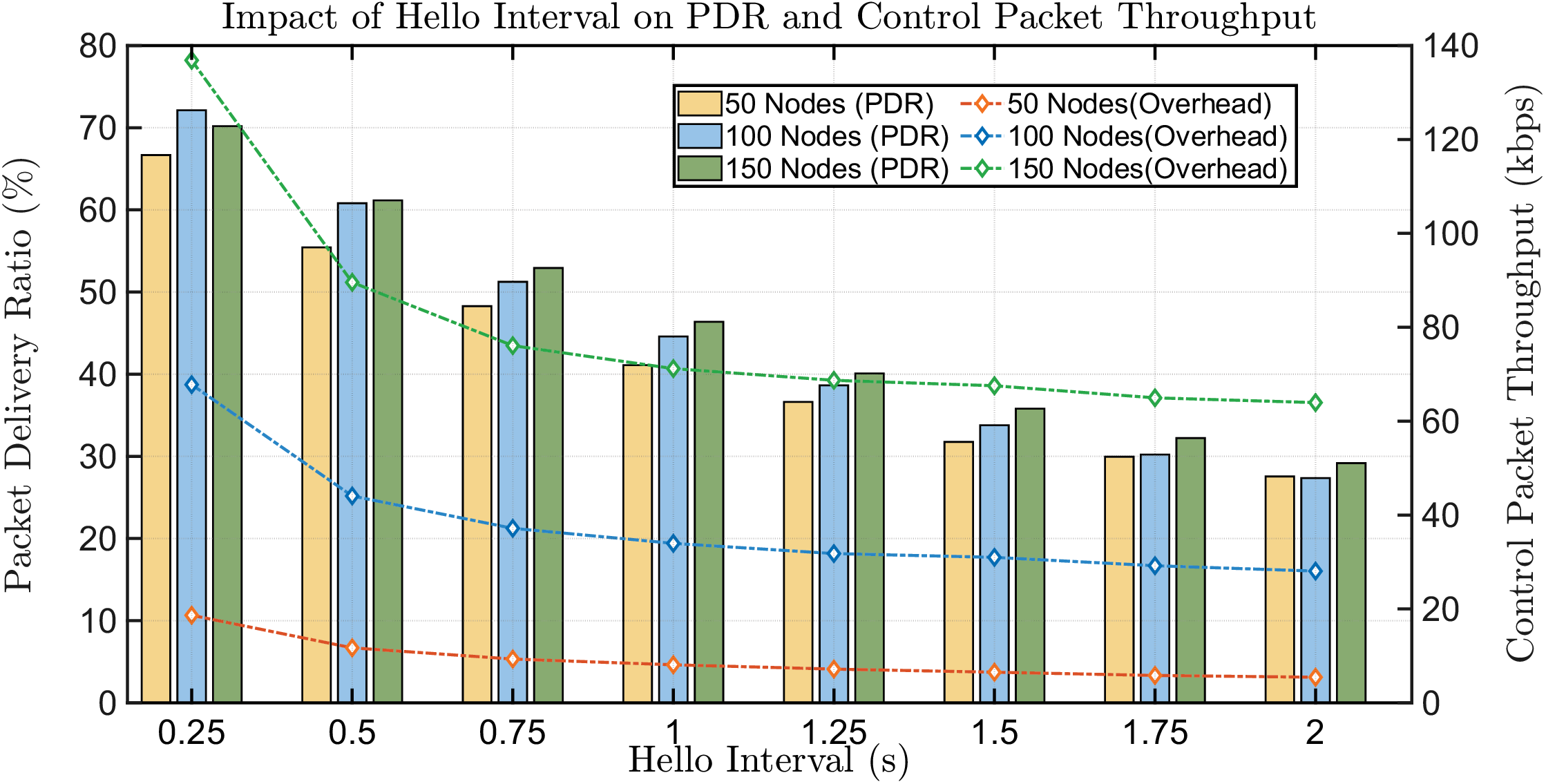}}
\caption{Packet delivery ratio and control packet throughput under different hello intervals of protocol OLSR.}
\label{fig:HelloForOLSR}
\end{figure}

\textbf{The balance between control overhead and the timeliness of topology awareness.} Fig.~\ref{fig:HelloForOLSR} shows that a shorter hello interval improves topology awareness but incurs higher control overhead. Multi-factor cluster-based algorithms require accurate real-time topology inputs, yet most existing protocols still rely on rigid, fixed-interval hello beaconing~\cite{icra,pica,MWCRSF}. Reinforcement learning~\cite{eeHello} and deep reinforcement learning~\cite{DRL-MLsA} have been applied to adapt the control interval dynamically, while they often struggle with convergence in large state spaces or assume limited network scales. To tackle this bottleneck, we introduce a lightweight fuzzy logic-based hello beacon optimizer to ensure accurate topology awareness by adaptively tuning the signaling frequency based on node mobility and density, thereby mitigating severe control overhead.

Overall, our contributions could be summarized as follows:
\begin{itemize}
   \item We propose a resilience evaluation framework that fuses multiple QoS dimensions into a composite score and quantifies protocol sensitivity to environmental variations. Analysis under both node scale and mobility variation scenarios shows that the three xBCR variants (BCR, QBCR and FBCR) consistently outperform other baseline schemes, with FBCR achieving the highest composite performance and all three ranking in the top three in overall resilience scoring.
   \item We design a dual-layer beacon maintenance mechanism for clusters to enable continuous bidirectional liveness monitoring. Simulation results show that the proposed scheme achieves competitive cluster lifetime, clustering ratio and low role transition frequency while avoiding frequent global re-clustering.
   \item We introduce a fuzzy logic-based hello interval optimizer that adaptively adjusts signaling frequency according to node mobility and neighbor density. Compared with fixed interval baselines, the proposed optimizer reduces control overhead by 25\% while maintaining comparable routing performance.
\end{itemize}

The remainder of this paper is organized as follows: Section~\ref{sec:related} reviews related works on clustering technologies, optimization algorithms and performance metrics. Section~\ref{sec:model} details the proposed framework, including the dual-layer cluster beacon mechanism, the hello interval adapter based on fuzzy logic, and the routing scheme. Section~\ref{sec:result} provides the simulation setup and discusses the results, highlighting the superiority of the proposed approach through the novel performance metrics. Finally, Section~\ref{sec:conclusion} concludes this paper.

\section{Related Works}\label{sec:related}

Clustering techniques can effectively simplify the link structure of large-scale networks and have become an important class of routing for FANETs~\cite{lakew2020routing}. Existing studies on cluster-based routing mainly focus on cluster-head election, cluster maintenance, and routing forwarding and optimization. To enhance cluster-head stability and overall network performance, cluster-head election mechanisms have evolved from simple single-factor strategies, such as distance~\cite{K-MORP} or energy level~\cite{IMRL}, to multi-factor comprehensive evaluation approaches. In such designs, the suitability of a node as a cluster head is typically assessed by jointly considering multiple attributes such as neighbor degree, residual energy, signal quality, positional centrality, and mobility similarity~\cite{icra,pica,RL-cluster,MGO_JAYA}. To cope with continuously varying links, cluster membership must be dynamically updated. While many studies rely on periodic reclustering to refresh cluster structures~\cite{K-MORP,FMORT,IABC}, others introduce more refined mechanisms, such as node joining and leaving, cluster merging and splitting, and backup cluster-head selection, to avoid unnecessary global reconfiguration~\cite{pica,MWCRSF,RL-cluster,icra}.

In recent years, machine learning and bio-inspired algorithms have been widely adopted in cluster formation and routing optimization due to their strong adaptive capabilities~\cite{computational_intell_review}. For example, Q-learning enables nodes to adaptively adjust routing or clustering strategies through interactions with the environment~\cite{icra,TARRAQ,QTAR,QRIFC}. However, as the number of decision factors increases, the state space expands rapidly, leading to slow convergence and limiting real-time applicability. Deep reinforcement learning (DRL) approaches, such as Deep Q-Networks (DQN) and other DRL variants, leverage function approximation via deep neural networks to accelerate convergence~\cite{DRL-MLsA,parouting,drl-ssr,DRLFR,IABC,MWCRSF}. Nevertheless, these methods typically depend on training processes and impose higher computational and energy requirements on UAV nodes. In parallel, bio-inspired algorithms have attracted significant attention owing to their distributed nature and inherent robustness. Inspired by collective behaviors observed in dragonfly~\cite{FMORT}, bee colony~\cite{IABC}, ant colony~\cite{Q-ANT} and mountain gazelles~\cite{MGO_JAYA}, these approaches employ population-based search mechanisms to optimize clustering and routing decisions. Beyond swarm intelligence, some studies further draw inspiration from the unique behaviors of individual organisms, such as physarum polycephalum~\cite{pica} and fire hawk~\cite{dcfh}, to improve routing efficiency and data transmission performance. Fuzzy logic has also played an important role in dynamic and complex aerial network scenarios, as it is well suited for handling uncertainty and imprecise information inherent in highly mobile environments, and has been applied to problems such as secure routing~\cite{FTSR,FUBA} and power control~\cite{airpro-fl}.

From the perspective of performance evaluation, most existing studies primarily consider general communication metrics, including throughput, end-to-end delay, jitter, packet delivery ratio, and control overhead. However, the high mobility and rapidly changing topology of FANETs necessitate more expressive metrics to characterize the robustness of routing protocols under dynamic network conditions, as these conventional metrics alone are often insufficient to clearly distinguish how effectively routing protocols adapt to complex aerial environments. To address this limitation, some cluster-based routing studies further introduce structural metrics, such as the number of clusters, cluster formation time, and cluster lifetime, to evaluate the efficiency and stability of the clustering process itself~\cite{icw,MWCRSF,MGO_JAYA,dcfh}. Nevertheless, these metrics still mainly characterize static performance aspects and fail to capture the resilience of routing protocols in maintaining stable performance under parameter perturbations, network scaling, and complex mobility patterns.

\section{Bio-inspired Clustering and Routing Protocol}\label{sec:model}

In this section, we present the Bio-inspired Clustering and Routing (BCR) protocol, which leverages an Artificial Bee Colony (ABC) algorithm for cluster formation, a dual-layer beaconing mechanism for lightweight maintenance, and a fuzzy logic optimizer for adaptive signaling.

\subsection{ABC-based Cluster Head Election and Cluster Formation}
The cluster head election in BCR is based on the Artificial Bee Colony (ABC) algorithm, which mimics the foraging behavior of honey bee swarms. Through iterative optimization, the algorithm selects $\mathcal{C} = \lfloor N/10 \rfloor$ nodes with the highest fitness values as cluster head candidates, where $N$ denotes the total number of nodes in the network.

\subsubsection{Fitness Function Design}
The fitness value of each node is computed as a weighted combination of four normalized factors:
\begin{equation}
Fit_i = w_{1} \cdot T^{lst}_i + w_2 \cdot \eta^{energy}_i + w_3 \cdot \eta^{degree}_i + w_4 \cdot \bar{d}_i
\label{eq:fitness}
\end{equation}

\noindent where the four factors are:

\begin{itemize}
\item \textbf{Link Stability Time}: This factor evaluates the average predicted link lifetime with all neighbors. For each neighbor $j$, $T^{lst}_{i,j}$ is computed based on the relative position and velocity:
\begin{equation}
T^{lst}_{i,j} = \begin{cases}
T_{max} & \text{if } a = 0 \text{ (relatively stationary)} \\
\min(t^+, T_{max}) & \text{otherwise}
\end{cases}
\end{equation}
where $t^+$ is the smallest positive root of $at^2 + bt + c = 0$ with $a = \|\Delta \mathbf{v}\|^2$, $b = 2 \langle \Delta \mathbf{p}, \Delta \mathbf{v} \rangle$, and $c = \|\Delta \mathbf{p}\|^2 - R^2$. Here $\Delta \mathbf{p} = \mathbf{p}_j - \mathbf{p}_i$ and $\Delta \mathbf{v} = \mathbf{v}_j - \mathbf{v}_i$ denote the relative position and velocity, and $R$ is the communication range. $T^{lst}_i$ is then normalized as:
\begin{equation}
T^{lst}_i = \min\left(1, \frac{\bar{T}^{lst}_i}{T_{ref}}\right)
\end{equation}
where $\bar{T}^{lst}_i = \frac{1}{|\mathcal{N}_i|}\sum_{j \in \mathcal{N}_i} T^{lst}_{i,j}$ is the average LST and $T_{ref}$ is a reference value for a long link life time.

\item \textbf{Energy Level}: The ratio of residual to initial energy is:
\begin{equation}
\eta_i^{energy} = \frac{E_i^{res}}{E_i^{init}}
\end{equation}
and this factor favors nodes with higher remaining energy for prolonged cluster head operation.

\item \textbf{Neighbor Degree}: The normalized neighbor degree is:
\begin{equation}
\eta^{degree}_i = \frac{|\mathcal{N}_i|}{N - 1}
\end{equation}
where $|\mathcal{N}_i|$ is the number of neighbors and $N$ is the total node count, preferring well-connected nodes.

\item \textbf{Average Neighbor Distance}: Measures average proximity to neighbors:
\begin{equation}
\bar{d}_i = \frac{1}{|\mathcal{N}_i|} \sum_{j \in \mathcal{N}_i} \frac{1}{1 + d_{ij}}
\end{equation}
where $d_{ij} = \|\mathbf{p}_i - \mathbf{p}_j\|$ is the Euclidean distance, favoring centrally located nodes.
\end{itemize}

\noindent The default weight configuration is $w_i=0.25 (i=1,2,3,4)$, providing equal importance to link stability, energy availability, connectivity, and spatial centrality.

\subsubsection{Cluster Formation Initialize}
In the initial stage, each node broadcasts \texttt{HELLO} messages to get the neighbors' information and compute fitness values accordingly.
Cluster heads are then elected through the Artificial Bee Colony algorithm.
\begin{itemize}
    \item \textbf{Employed Bee} performs neighborhood search using $x^{'}_{i} = x_i + \phi(x_i - x_j)$ where $\phi \in [-1, 1]$;
    \item \textbf{Onlooker Bee} exploits promising solutions based on fitness proportional selection probability $P_i={Fit_i / \sum_{i=1}^{N} Fit_{i}}$;
    \item \textbf{Scout Bee} reinitializes stagnant solutions when $trial_i \geq 4N$ to escape local optima and maintain population diversity.
\end{itemize}

Following the ABC-based cluster head (CH) election process, $\mathcal{C}$ candidate cluster heads are identified and temporarily designated as “BackupHead”, while all other nodes are initialized to the “Idle” state. To recruit members, candidate CHs broadcast \texttt{CH\_CALL} messages with a transmission priority determined by their fitness values. Upon receiving the first \texttt{CH\_CALL}, a potential member node $CM_i$ initiates an observation window ($t=1\mathrm{s}$). During this interval, if $CM_i$ detects another \texttt{CH\_CALL} from $CH_j$ with superior fitness, $CM_i$ updates its preference to $CH_j$. After the window expires, $CM_i$ sends a \texttt{JOIN\_REQ} to the selected CH. Upon receiving a \texttt{JOIN\_REQ}, $CH_i$ replies with a \texttt{JOIN\_RESP} to confirm membership, and the association is finalized once $CM_i$ receives a \texttt{JOIN\_RESP}, at which point it transitions to the “Member” state.

\begin{figure}[htbp]
    \centering
    \includegraphics[width=\linewidth]{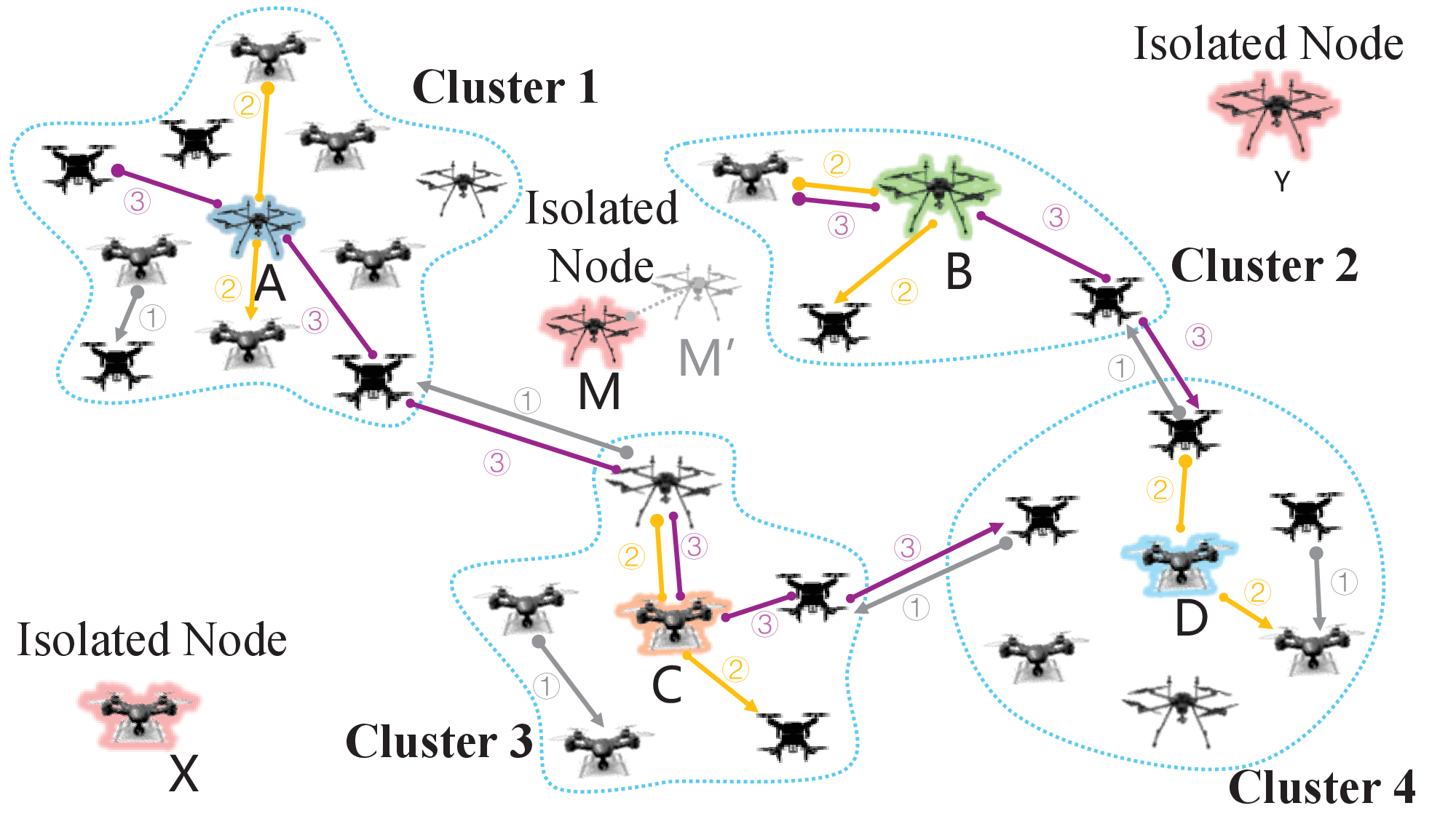}
    \caption{Cluster-based routing. Nodes A, B, C, and D act as the cluster heads of clusters 1, 2, 3, and 4, respectively. Gray links denote direct data transmissions between one-hop neighbors, yellow links represent intra-cluster data forwarding via the cluster head, and purple links indicate inter-cluster data forwarding through cluster heads.}
    \label{fig:clusteredRouting}
\end{figure}

Simulation observations indicate that some nodes may not immediately join a cluster following the initial phase, as illustrated in Fig.~\ref{fig:clusteredRouting}. This outcome is expected in high-mobility and sparsely distributed aerial scenarios. Specifically, nodes $X$ and $Y$ become isolated because they lack neighbors during the election phase, whereas node $M$ remains unclustered due to the absence of a designated CH within its communication range. In our scheme, nodes $X$ and $Y$ transition to form singleton clusters. Meanwhile, node $M$ enters a listening state for a predefined period to receive potential topology updates. If no suitable cluster becomes available before the timeout expires, node $M$ also establishes a singleton cluster to maintain network structure.

\subsection{Dual-Layer Beacon Cluster Maintenance Mechanism}

After initial cluster formation, BCR employs a Dual-Layer Beacon mechanism to maintain cluster structure through bidirectional liveness monitoring.

\subsubsection{Beacon Types}
Cluster heads broadcast \texttt{CH\_LIVE} every $T_{beacon}$~s, containing node ID, cluster ID, position, velocity, and fitness value. Cluster members unicast \texttt{CM\_LIVE} to their cluster head upon receiving beacon from the cluster head, confirming continued participation.

\subsubsection{Timeout Detection}
Both cluster heads and members implement timeout detection with threshold $\tau = 5 T_{beacon}$:

\begin{itemize}
\item \textbf{Cluster Head:} Removes members exceeding the timeout and resets their role to “Idle”.
\item \textbf{Cluster Member:} Leaves the cluster if the cluster head times out or disappears from the neighbor table, then initiates rejoin.
\end{itemize}

\begin{algorithm}[ht]
\caption{Dual-Layer Beacon Maintenance for Cluster}
\label{alg:beacon}
\begin{algorithmic}[1]
\STATE \textbf{Input:} $T_{beacon} = 2.5s$

\STATE \textbf{Layer 1: Cluster Head Beacon (\texttt{CH\_LIVE})}
\IF{$Role == $“Head”}
    \STATE Broadcast \texttt{CH\_LIVE}
    \STATE Collect \texttt{CM\_LIVE} responses from members
    \FOR{each member $m$ in cluster}
        \IF{$t_{now} - Member.LastUpdate > \tau$}
            \STATE Remove $m$ from cluster \COMMENT{Member timeout}
        \ENDIF
    \ENDFOR
    \IF{$Member.size <=1$ and $ValidNeighborNum > 0$}
        \STATE RemoveCluster() and Set $Role \gets $“Idle”
        \STATE Join to other cluster
    \ENDIF
\ENDIF

\STATE \textbf{Layer 2: Cluster Member Beacon (\texttt{CM\_LIVE})}
\IF{$Role$ == “Member”}
    \IF{CH\_LIVE received from cluster head}
        \STATE Unicast \texttt{CM\_LIVE} to CH
        \STATE Reset timeout counter
    \ELSIF{$t_{now} - Head.LastUpdate > \tau$}
        \STATE Set $Role \gets $ "Isolate"
        \STATE Trigger rejoin \COMMENT{CH timeout}
    \ENDIF
\ENDIF

\STATE Schedule next beacon cycle after $T_{beacon}$
\end{algorithmic}
\end{algorithm}

When a cluster head detects all members have departed, it evaluates neighbor availability: if valid neighbors exist, the cluster dissolves and the node rejoins another cluster; otherwise, it maintains isolated cluster head status. The maintenance procedure is detailed in Algorithm~\ref{alg:beacon}.

%------------------------------------------------------------------------------
\subsection{Clustering Based Routing Algorithm}
%------------------------------------------------------------------------------

After cluster formation, nodes forward data packets by jointly leveraging neighbor information and cluster structure, following a priority-based strategy as shown in Fig.~\ref{fig:clusteredRouting} and Algorithm~\ref{alg:routing}.

If the destination is a one-hop neighbor (gray links in Fig.~\ref{fig:clusteredRouting}), packets are forwarded directly to minimize delay. For destinations within the same cluster (yellow links), member nodes relay packets to the cluster head, which then delivers them directly or via greedy geographic forwarding. When communicating across clusters (purple links), member nodes forward packets to their associated cluster head, which is responsible for inter-cluster delivery to improve scalability.

To cope with dynamic topology changes, greedy geographic forwarding is adopted as a fallback mechanism when valid cluster information is unavailable, ensuring continuous packet delivery under highly dynamic conditions.

\begin{algorithm}[htbp]
\caption{Clustering-based Routing Algorithm}
\label{alg:routing}
\begin{algorithmic}[1]
\REQUIRE Packet $p$, destination $dst$
\ENSURE Next hop address
\IF{$dst$ is broadcast}
    \RETURN BroadcastRoute($dst$)
\ENDIF
\STATE \textbf{// Strategy 1: Direct neighbor}
\IF{IsNeighbor($dst$)}
    \RETURN $dst$
    \ELSE 
        \RETURN GetClusterHeadAddress()
\ENDIF
\STATE \textbf{// Strategy 2: Intra-cluster}
\IF{$ClusterId == dstClusterId$}
    \IF{$Role \neq$ “Head”}
        \RETURN GetClusterHeadAddress()
    \ELSE
        \RETURN GreedyForward($dst$)
    \ENDIF
\ENDIF
\STATE \textbf{// Strategy 3: Inter-cluster}
\IF{$Role \neq$ “Head”}
    \RETURN GetClusterHeadAddress()
\ENDIF
\STATE \textbf{// Strategy 4: Greedy fallback}
\RETURN GreedyNeighbor($\mathbf{p}_{dst}$, $\mathbf{p}$, $prevHop$)
\end{algorithmic}
\end{algorithm}

\subsection{Fuzzy Logic Based Hello Beacon Optimizer}
As mentioned in Section~\ref{sec:Intro}, fixed interval cannot adapt to varying network dynamics: too short intervals waste energy and cause channel congestion, while too long intervals lead to stale neighbor information and degraded PDR. To address this challenge, we propose a fuzzy logic based BCR (FBCR) that dynamically adjusts the hello beacon interval based on real-time network conditions. Q-learning~\cite{DRL-MLsA} and fuzzy logic~\cite{airpro-fl} have been applied to hello interval control in prior work. 

However, as mentioned in Section~\ref{sec:related}, reinforcement learning algorithms may encounter problems such as excessively large state spaces and slow convergence in large-scale network scenarios. Meanwhile, the UAV deployment scale considered in~\cite{airpro-fl} is limited, and the node mobility patterns are relatively regular. In contrast, the scenario addressed in this paper involves a significantly larger number of UAVs with irregular and highly dynamic mobility, under which adaptive beacon interval control becomes considerably more challenging.

\subsubsection{Fuzzification}

FBCR employs three input variables to characterize network dynamics:

\textbf{Relative Velocity ($v_r$):} The average relative velocity between the local node $i$ and its neighbors:
\begin{equation}
v_r = \frac{1}{|\mathcal{N}|} \sum_{j \in \mathcal{N}} \|\mathbf{v}_j - \mathbf{v}_i\|
\end{equation}

\textbf{Neighbor Density ($\rho$):} The normalized neighbor count is defined as $\rho = |\mathcal{N}| / N_{\max}$, where $N_{\max}$ is a reference value representing "dense", and is set to 80 in simulation.

\textbf{Link Stability ($\epsilon$):} The minimum predicted link lifetime among all neighbors:
\begin{equation}
\epsilon = \min_{j \in \mathcal{N}} \epsilon_j, \quad \epsilon_j = \frac{R_{max} - d_{ij}}{v_{radial}^{(j)}}
\end{equation}

\noindent where $d_{ij}$ is the distance to neighbor $j$, and $v_{radial}^{(j)}$ is the radial velocity component (positive when moving apart).

%-------------------------fig-fuzzy-optimizer-----------
\begin{figure}[htbp]
\centerline{\includegraphics[width=\linewidth]{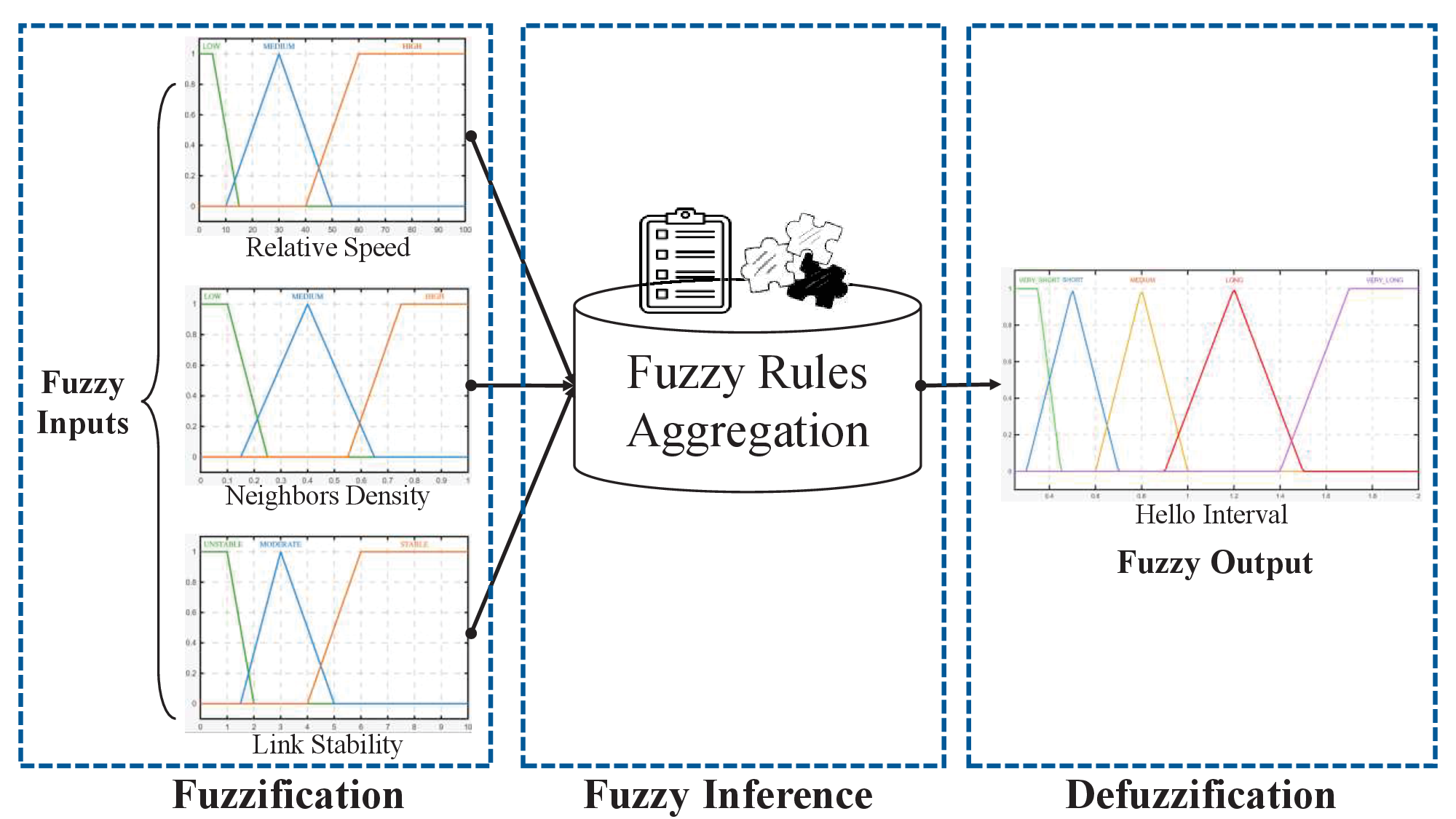}}
\caption{Fuzzy hello optimizer.}
\label{fig:fuzzy}
\end{figure}

Each input variable is associated with three linguistic terms (\texttt{LOW}/\texttt{MEDIUM}/\texttt{HIGH} or \texttt{UNSTABLE}/\texttt{MODERATE}/\texttt{STABLE}), and the output variable hello interval $T_h \in [0.25, 2.0]$s has five terms (\texttt{VERY\_SHORT}, \texttt{SHORT}, \texttt{MEDIUM}, \texttt{LONG}, \texttt{VERY\_LONG}). The membership functions are illustrated in Fig.~\ref{fig:fuzzy}.%.~\ref{fig:input_memberships} and Fig.~\ref{fig:output_membership}.

%------------------------table-25 IF_THEN logic-----------------
\begin{table}[htbp]
    \centering
    \caption{Optimized fuzzy rules base for hello beacon interval}
    \label{tab:rules}
    \renewcommand{\arraystretch}{1.2}
    \begin{tabular}{c|c|ccc}
        \toprule
        \multirow{2}{*}{Density $\rho$} & \multirow{2}{*}{Rel. Speed $v_r$} & \multicolumn{3}{c}{Link Stability $\epsilon$} \\
        \cmidrule{3-5}
         & & \texttt{UNSTABLE} & \texttt{MODERATE} & \texttt{STABLE} \\
        \midrule
        \multirow{3}{*}{\texttt{LOW}}
         & \texttt{LOW}    & S  & M  & VL \\
         & \texttt{MEDIUM} & S  & M  & L  \\
         & \texttt{HIGH}   & VS & S  & M  \\
        \midrule
        \multirow{3}{*}{\texttt{MEDIUM}}
         & \texttt{LOW}    & S  & L  & VL \\
         & \texttt{MEDIUM} & S  & M  & L  \\
         & \texttt{HIGH}   & VS & S  & M  \\
        \midrule
        \multirow{3}{*}{\texttt{HIGH}}
         & \texttt{LOW}    & S  & L  & L  \\
         & \texttt{MEDIUM} & S  & M  & L  \\
         & \texttt{HIGH}   & VS & S  & M  \\
        \bottomrule
        \multicolumn{5}{l}{\footnotesize \textit{VS=Very Short, S=Short, M=Medium, L=Long, VL=Very Long}}
    \end{tabular}
\end{table}

\subsubsection{The Inference Engine}
FBCR utilizes a Mamdani-type inference engine to adapt the hello interval. We have established 25 IF-THEN rules, as structured in Table~\ref{tab:rules}, to regulate the hello interval. The fuzzy inference process is detailed as follows:

Specifically, the inference process consists of rule evaluation and aggregation. In the rule evaluation phase, the firing strength $\alpha_k$ for the $k$-th rule is computed using the \textit{minimum operator}:
\begin{equation}
\alpha_k = \min(\mu_{A_{k,1}}(\rho), \mu_{A_{k,2}}(v_r), \mu_{A_{k,3}}(\epsilon))
\end{equation}
where $\mu_{A_{k,j}}$ denote the membership degrees for density $\rho$, relative speed $v_r$, and link stability $\epsilon$, respectively. Subsequently, the aggregation step utilizes the \textit{maximum operator} to determine the final membership $\mu_i$ for each output level $B_i$:
\begin{equation}
\mu_{i} = \max_{k: B_k = B_i} \alpha_k, \quad i \in \{VS, S, M, L, VL\}
\end{equation}

\subsubsection{Defuzzification}
To obtain a crisp control value, the Center of Gravity (COG) method is applied. We discretize the output space into five centers $c_i \in \{0.35, 0.50, 0.80, 1.20, 1.70\}$ (in seconds). The optimized interval $T_h^*$ is calculated as:
\begin{equation}
T_h^* = \frac{\sum_{i=1}^{5} c_i \cdot \mu_i}{\sum_{i=1}^{5} \mu_i}
\end{equation}

Finally, $T_h^*$ is clamped to $[0.25, 2.0]$ seconds to ensure protocol stability and comply with FANET physical constraints.

\subsection{Parameter Sensitivity Index and Balance Index Definition}

To differentiate the sensitivity of routing schemes under dynamic aerial networking conditions, we introduce a composite performance index (CPI) that integrates packet delivery ratio, end-to-end delay, throughput, and control overhead:
\begin{equation}
CPI = \sum_{i=1}^{n} w_i \cdot f_i(\tilde{x}_i),
\end{equation}
where $\sum_i w_i = 1$ and each metric is min-max normalized to $[0,1]$ over the joint set of all (protocol, parameter) combinations in the sweep, $\tilde{x}_i = (x_i - x_i^{\min})/(x_i^{\max} - x_i^{\min})$, so that CPI values remain directly comparable across protocols. The mapping $f_i$ takes $\tilde{x}_i$ when metric $i$ is ``higher is better'' and $1-\tilde{x}_i$ otherwise. Unless stated otherwise, we use $n=4$ with equal weights $w_i = 0.25$.

Based on CPI, the Parameter Sensitivity Index (PSI) characterizes robustness with respect to an environment parameter
$P_m \in \mathbb{P}$ (e.g., network scale or node mobility). Let $\{p_j\}_{j=1}^{|P_m|}$ denote its sampled values in strictly increasing order. Define
\begin{equation}
\mathrm{PSI}_m = |\mathrm{MPS}_m|\,(1 + \sigma_m)\,(1 - R^2_m),
\end{equation}
where
\begin{align}
\mathrm{MPS}_m &= \frac{1}{|P_m|-1}
  \sum_{i=1}^{|P_m|-1}
  \frac{CPI(p_{i+1}) - CPI(p_i)}{p_{i+1} - p_i}, \\
\sigma_m &= \mathrm{std}\!\left(
  \frac{CPI(p_{i+1}) - CPI(p_i)}{p_{i+1} - p_i}
  \right)_{i=1}^{|P_m|-1}, \\
R^2_m &= 1 - \frac{\sum_{j=1}^{|P_m|} (CPI_j - \hat{CPI}_j)^2}
                  {\sum_{j=1}^{|P_m|} (CPI_j - \overline{CPI})^2},
\end{align}
with $\hat{CPI}_j$ the value at $p_j$ from an ordinary least-squares linear fit of $CPI$ on $p$, and $\overline{CPI} = \frac{1}{|P_m|} \sum_j CPI(p_j)$. Intervals with $p_{i+1}=p_i$ are excluded, and we set $\mathrm{PSI}_m \triangleq 0$ when $\mathrm{MPS}_m = 0$. Lower $\mathrm{PSI}_m$ indicates more robust behavior.

To jointly characterize performance and robustness across the $Q$ compared protocols, each protocol is mapped to a point in the performance--resilience plane. The PSI is normalized across protocols and converted to a stability score:
\begin{equation}
\widetilde{\mathrm{PSI}}_q
  = \frac{\mathrm{PSI}_q}{\max_{1\le i\le Q}\mathrm{PSI}_i},
\qquad \mu_q = 1 - \widetilde{\mathrm{PSI}}_q \in [0,1],
\end{equation}
yielding the balance index $\beta$ as a convex combination of performance and resilience:
\begin{equation}
\mathrm{\beta}_q = w_p\,\overline{CPI}_q + w_s\,\mu_q,
\quad w_p + w_s = 1,
\label{eq:bi}
\end{equation}
where the specific choice of $(w_p, w_s)$ is given in Section~\ref{sec:result}.

\section{Simulation and Result Evaluation}\label{sec:result}

This section presents the simulation setup and performance evaluation conducted to validate the effectiveness of the proposed algorithms and models. Considering that simulations with more than 200 nodes incur prohibitively long execution times, identical simulation environments were deployed on multiple hosts, all configured with Ubuntu 22.04 and NS-3.40. The detailed simulation parameters are summarized in Table~\ref{tab:simulation_parameters}.

\begin{figure*}[htbp]
\centering
% --- 左侧栏的图 ---
\begin{minipage}[b]{0.48\linewidth}
\centering
\includegraphics[width=\linewidth]{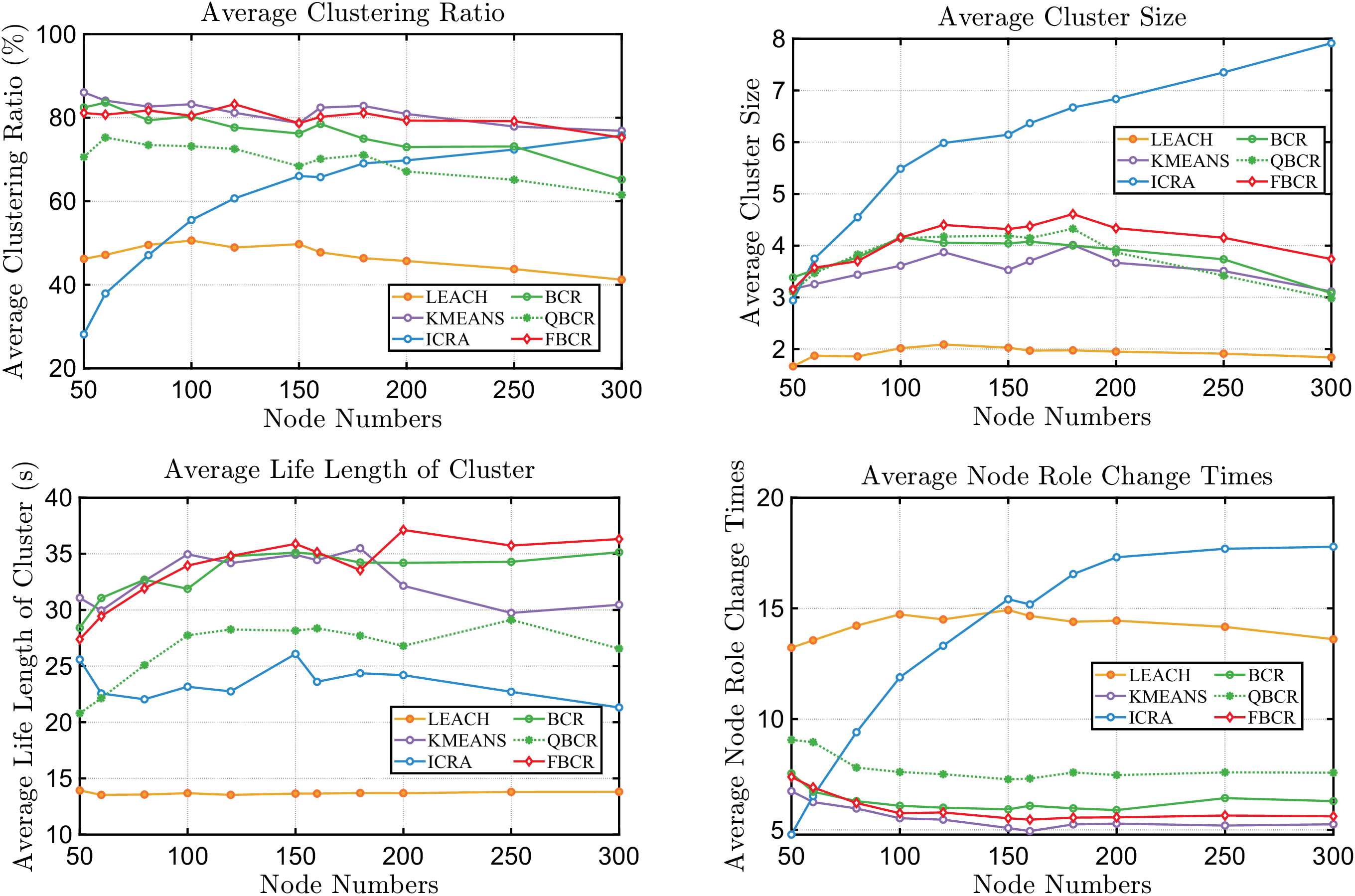}
\caption{Impact of node scale on clustering.}
\label{fig:clusteringCompare}
\end{minipage}
\hfill % 弹性间距，把左右两张图推开
% --- 右侧栏的图 ---
\begin{minipage}[b]{0.48\linewidth}
\centering
\includegraphics[width=\linewidth]{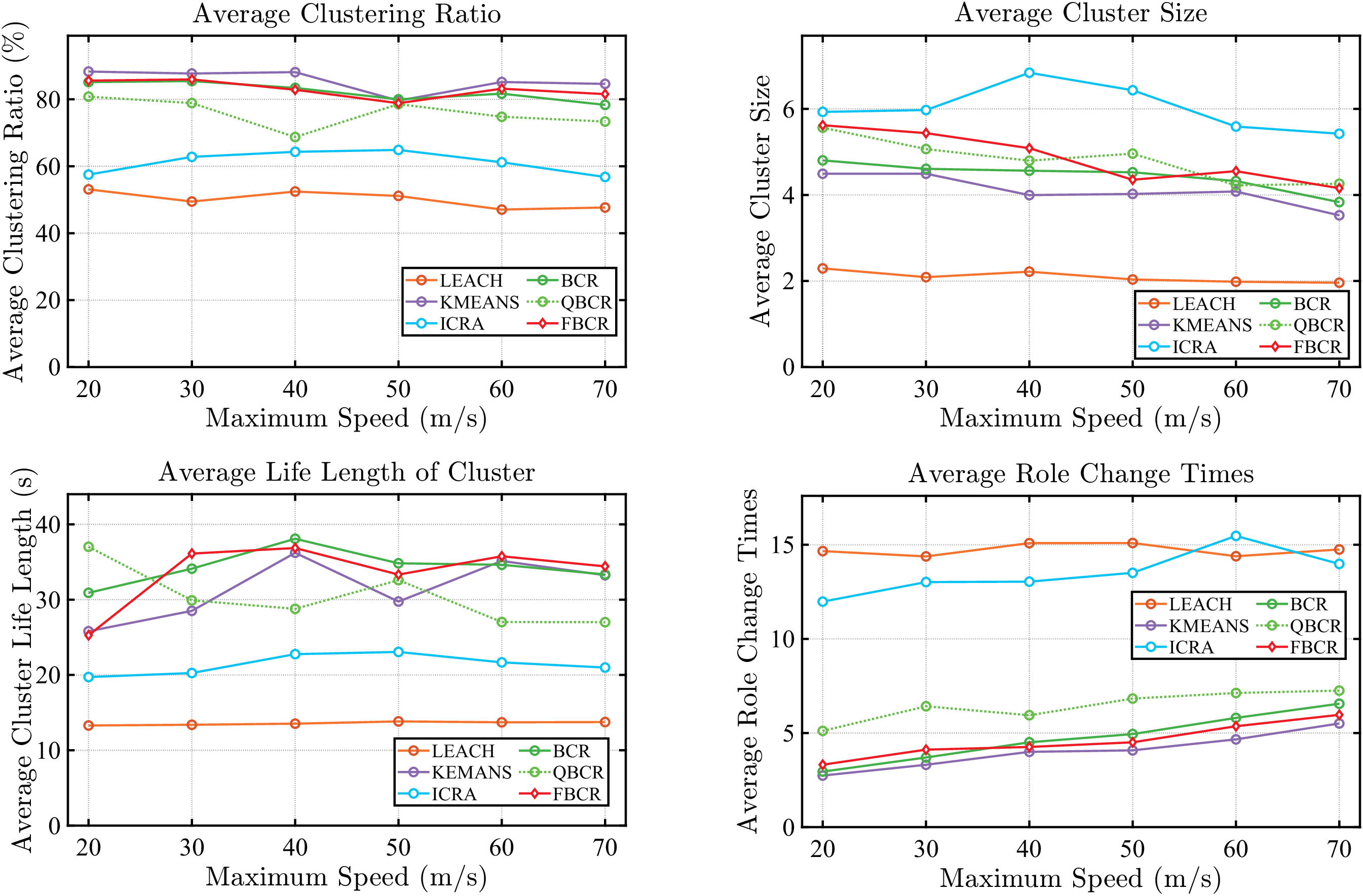}
\caption{Impact of node speed on clustering.}
\label{fig:clusteringCompareSpeed}
\end{minipage}

\end{figure*}

% ----------------------QoS_Scale-------------------------------
\begin{figure*}[!t]
\centering
\subfloat[]{\includegraphics[width=2.25in]{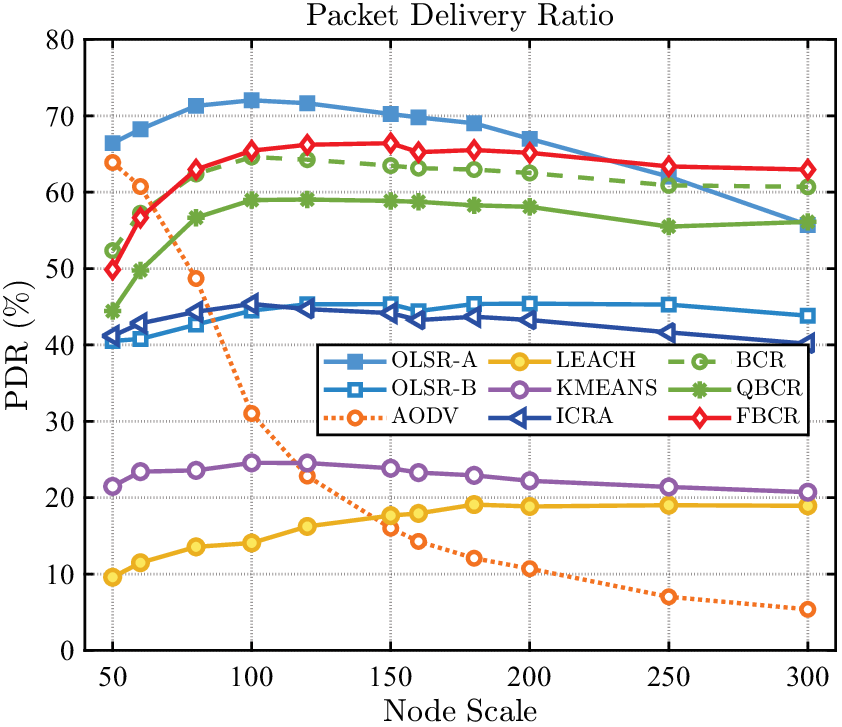}%
\label{fig:Scale_PDR}}
\hfil
\subfloat[]{\includegraphics[width=2.25in]{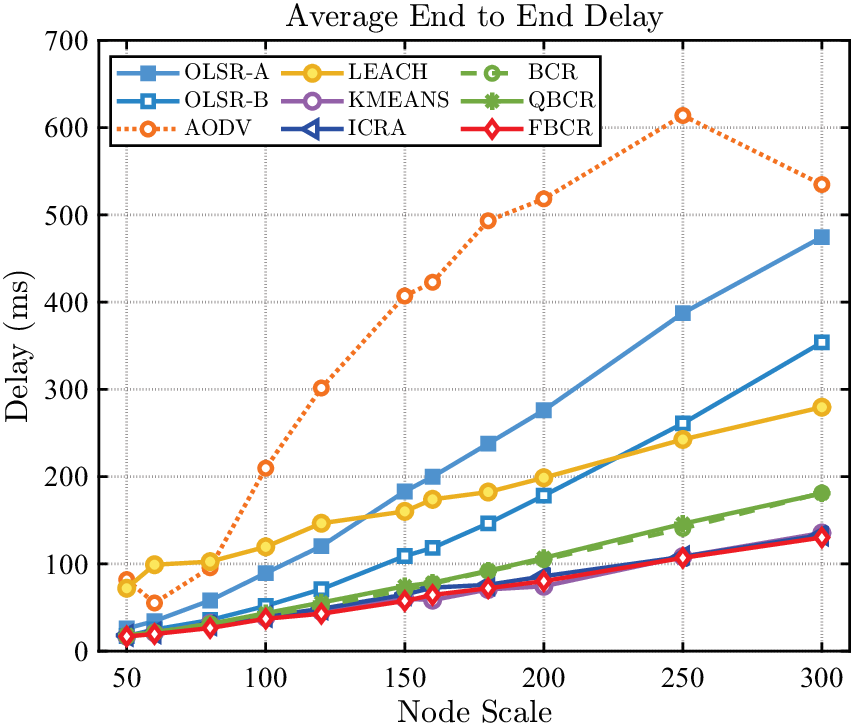}%
\label{fig:Scale_E2ED}}

\subfloat[]{\includegraphics[width=2.25in]{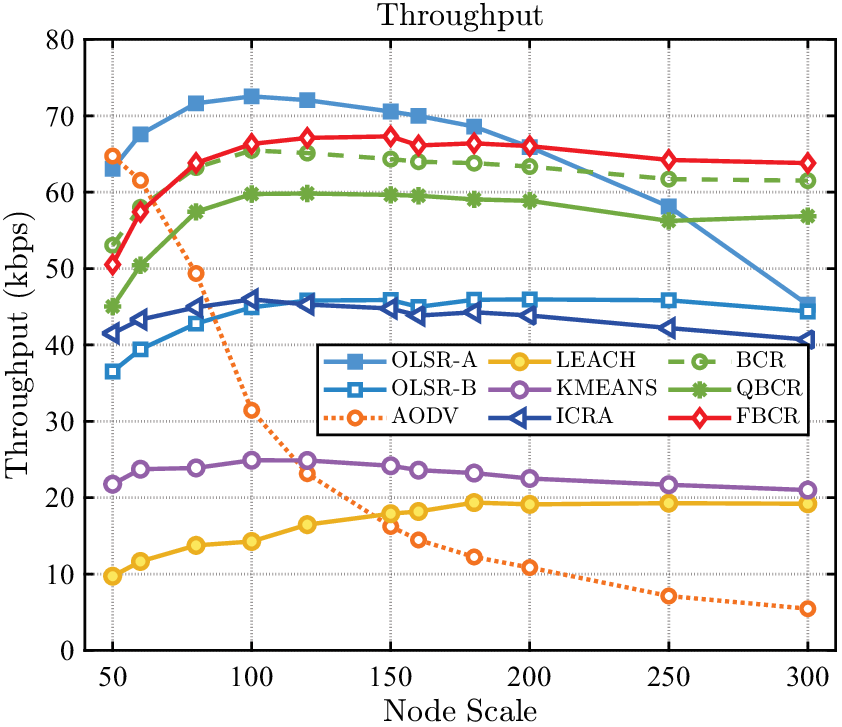}%
\label{fig:Scale_THP}}
\hfil
\subfloat[]{\includegraphics[width=2.25in]{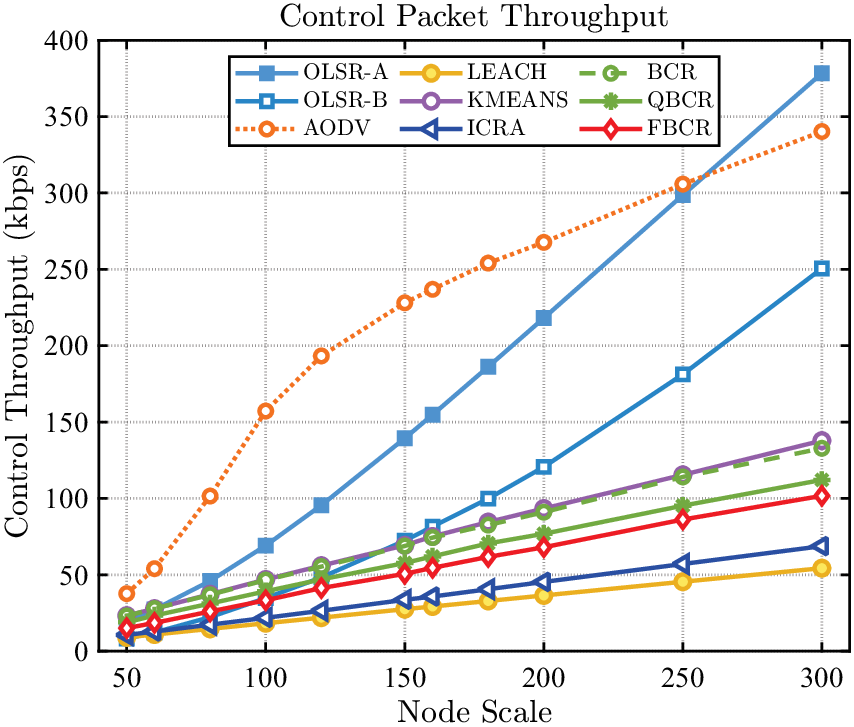}%
\label{fig:Scale_CTL_THP}}

\caption{Impact of node scale on QoS. (a) packet delivery ratio. (b) average end to end delay. (c) network throughput. (d) control packet throughput.}
\label{fig:4_in_one_scale}
\end{figure*}

%-----------------------Table of Simulation parameters
\begin{table}[htbp]
\centering
\caption{Simulation Parameters}
\label{tab:simulation_parameters}
\begin{tabular}{ll}
\hline
\textbf{Parameter} & \textbf{Value} \\
\hline
Network Area & $2000 \times 2000 \times 500$ m$^{3}$ \\
Number of UAVs & 50--300 \\
Mobility Model & Gauss--Markov \\
Node Speed & 20--60 m/s \\
Communication Range & 500 m \\
MAC Protocol & IEEE 802.11g \\
Traffic Type & UDP \\
Testflow(src,dst)  & 100 Pairs\\
Packet Size & 100 bytes \\
Simulation Duration & 200 s \\
\hline
\end{tabular}
\end{table}

To investigate the impact of network scale and node mobility on routing performance, two categories of simulation scenarios were considered. In the first category, the node speed range was fixed at 20--60~m/s, while the number of nodes varied across 50, 60, 80, 100, 120, 150, 160, 180, 200, 250, and 300. In the second category, the number of nodes was fixed at 150 and the minimum node speed was set to 10~m/s, whereas the maximum speed was varied among 20, 30, 40, 50, 60, and 70~m/s. For each scenario, routing performance metrics were collected over multiple random seeds, and the reported results correspond to the averaged values.

To demonstrate the efficiency of the proposed scheme, two classical ad hoc routing protocols, AODV and OLSR, were selected as baselines. For OLSR, two configurations were evaluated with hello intervals set to 0.25~s (denoted as OLSR-A) and 1~s (denoted as OLSR-B), respectively. In terms of cluster-based routing, two classical clustering algorithms, LEACH and K-means, as well as a state-of-the-art scheme, ICRA~\cite{icra}, were included for comparison. Furthermore, to assess the impact of different hello interval optimization strategies, three variants of BCR were evaluated: BCR with a fixed hello interval, QBCR employing Q-learning-based hello interval optimization, and the proposed FBCR, which integrates the proposed fuzzy logic-based hello interval optimizer.

%-------------------QoS_Speed---------------------
\begin{figure*}[!t]
\centering
\subfloat[]{\includegraphics[width=2.25in]{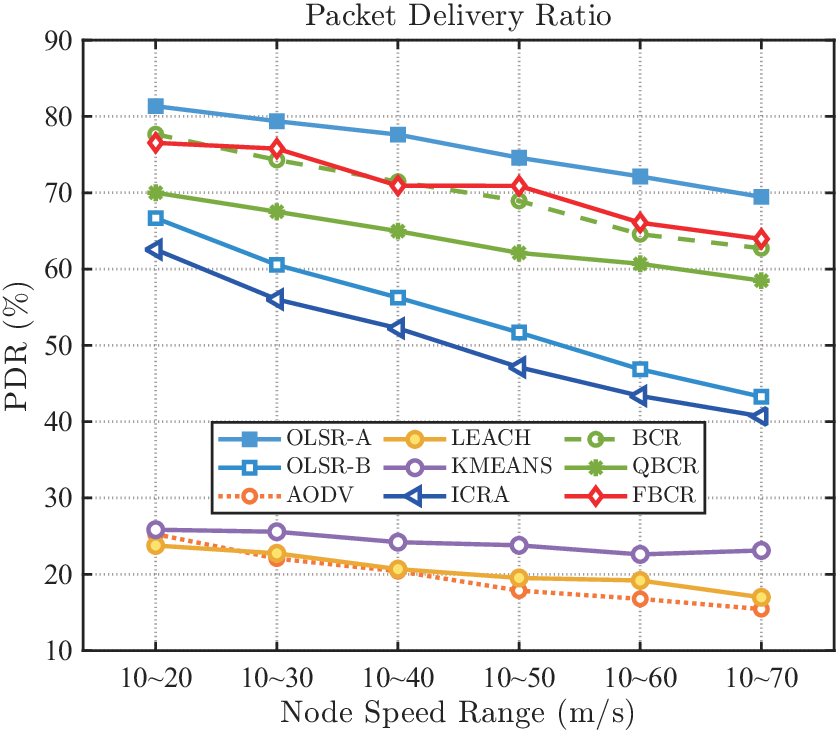}%
\label{fig:Speed_PDR}}
\hfil
\subfloat[]{\includegraphics[width=2.25in]{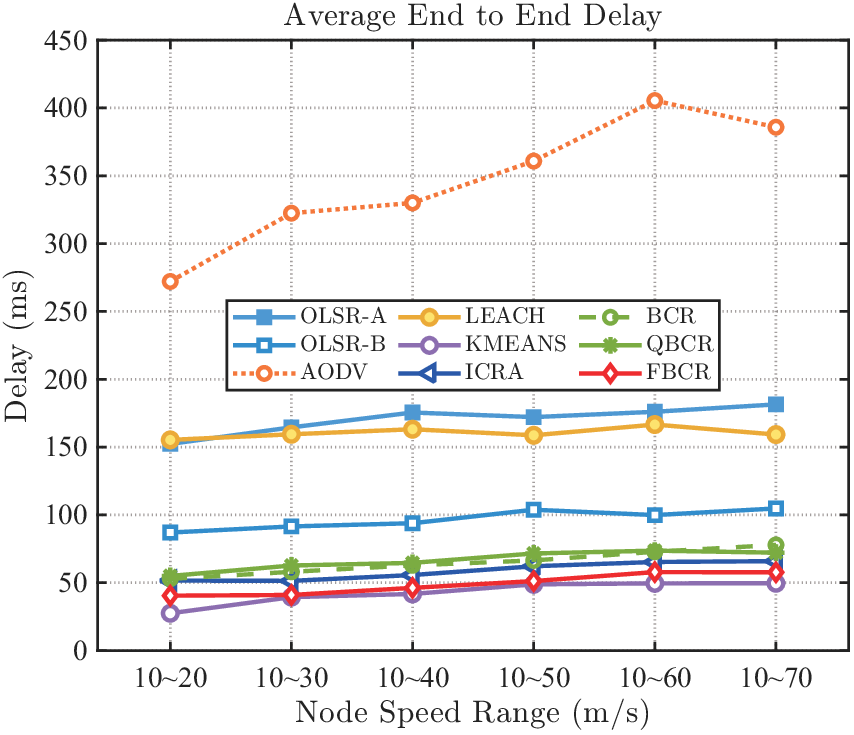}%
\label{fig:Speed_E2ED}}

\subfloat[]{\includegraphics[width=2.25in]{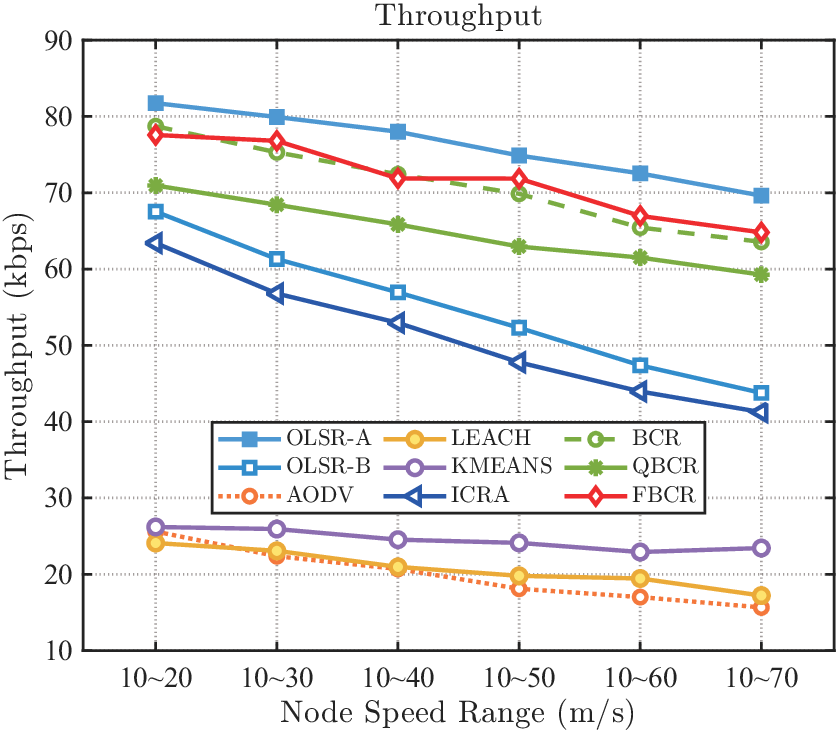}%
\label{fig:Speed_THP}}
\hfil
\subfloat[]{\includegraphics[width=2.25in]{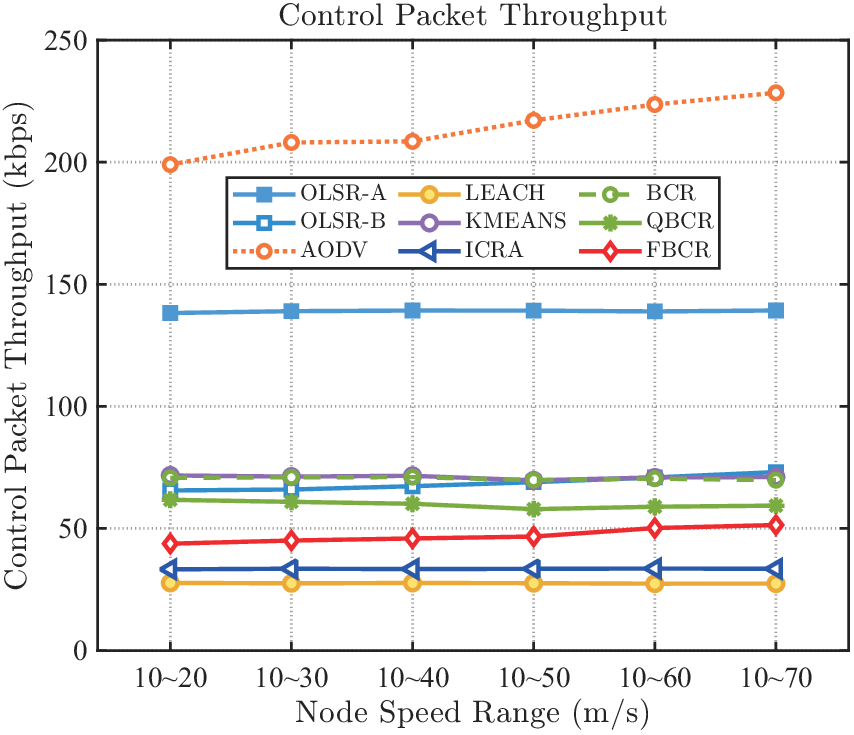}%
\label{fig:Speed_CTL_THP}}

\caption{Impact of node speed on QoS. (a) packet delivery ratio. (b) average end to end delay. (c) network throughput. (d) control packet throughput.}
\label{fig:4_in_one_speed}
\end{figure*}

% \begin{figure}[htbp]
% \centerline{\includegraphics[width=9cm]{figs/clustering_compare.eps}}
% \caption{Impact of node scale on clustering.}
% \label{fig:clusteringCompare}
% \end{figure}

% \begin{figure}[htbp]
% \centerline{\includegraphics[width=9cm]{figs/clustering_compare_speed.eps}}
% \caption{Impact of speed on clustering.}
% \label{fig:clusteringCompareSpeed}
% \end{figure}

\subsection{Clustering efficiency Comparison}

The proposed algorithm, together with K-means, LEACH, and ICRA, can be categorized as cluster-based routing schemes. Clustering-related information of all nodes (e.g., roles and associated cluster IDs) was collected every 5~s starting from $t = 10$~s during the simulation. Several clustering metrics were compared in the simulation, including the average cluster participation ratio ($\bar{\eta}=\frac{1}{N} \sum^{K}_{i=1}|c_i|$), average cluster size ($\bar{\lambda} = \frac{1}{K} \sum^{j}_{i=1}|c_i|$), average cluster lifetime ( $\bar{T_c} =\frac{1}{h} \sum_{i=1}^{h}\Delta T_i $), and the average number of role transitions per node ($\bar{\delta} =\frac{1}{N} \sum_{i=1}^{N}\delta_{i}$), where $K$ is the number of valid clusters, $h$ denotes the total number of clusters over the simulation, $|c_i|$ is the size of cluster $i$, $\Delta T_i$ is its effective lifetime, and $\delta_i$ denotes the role change times of node $i$.

Fig.~\ref{fig:clusteringCompare} and~\ref{fig:clusteringCompareSpeed} illustrate the clustering efficiency under different network scale and node speed ranges. It can be observed that the proposed FBCR consistently achieves competitive performance across all evaluated clustering metrics and ranks among the leading schemes in both scenarios. In contrast, LEACH exhibits suboptimal performance across all metrics, indicating its limited adaptability in highly dynamic FANET environments.

\subsection{Routing and Packet Forwarding Performance Comparison}

In this section, four commonly used QoS metrics are evaluated, including packet delivery ratio (PDR), average end-to-end delay (E2ED), throughput, and control packet throughput. For each metric, the final results are obtained by averaging the outcomes over eight independent runs with different random seeds.

Fig.~\ref{fig:4_in_one_scale} and~\ref{fig:4_in_one_speed} present the QoS performance across different scenarios. As shown in Fig.~\ref{fig:4_in_one_scale}, the PDR and throughput of AODV degrade noticeably as the network size increases, while its E2ED and control overhead remain consistently high. By reducing the hello interval, OLSR-A achieves the best PDR and throughput when the number of nodes is no greater than 200; however, when the network size exceeds 250 nodes, its performance is surpassed by the proposed scheme due to escalating control overhead. Cluster-based schemes exhibit relatively moderate performance variations as the network scales.
The three xBCR variants outperform the baseline schemes in PDR, E2ED, and throughput, with FBCR being particularly notable: it maintains comparable routing performance to BCR while reducing control overhead by 25\%.

As illustrated in Fig.~\ref{fig:4_in_one_speed}, the trends of PDR and throughput with respect to node speed are generally similar across schemes. Compared with network scale, node speed has a relatively smaller impact on E2ED and control overhead. Although OLSR-A consistently achieves the highest PDR and throughput, its E2ED and control overhead are inferior to all other schemes except AODV. The proposed FBCR maintains balanced performance and consistently ranks among the top three across all metrics.

\subsection{Comprehensive Performance and Resilience Analysis}

\begin{figure*}[!t]
\centering
\subfloat[]{\includegraphics[width=2in]{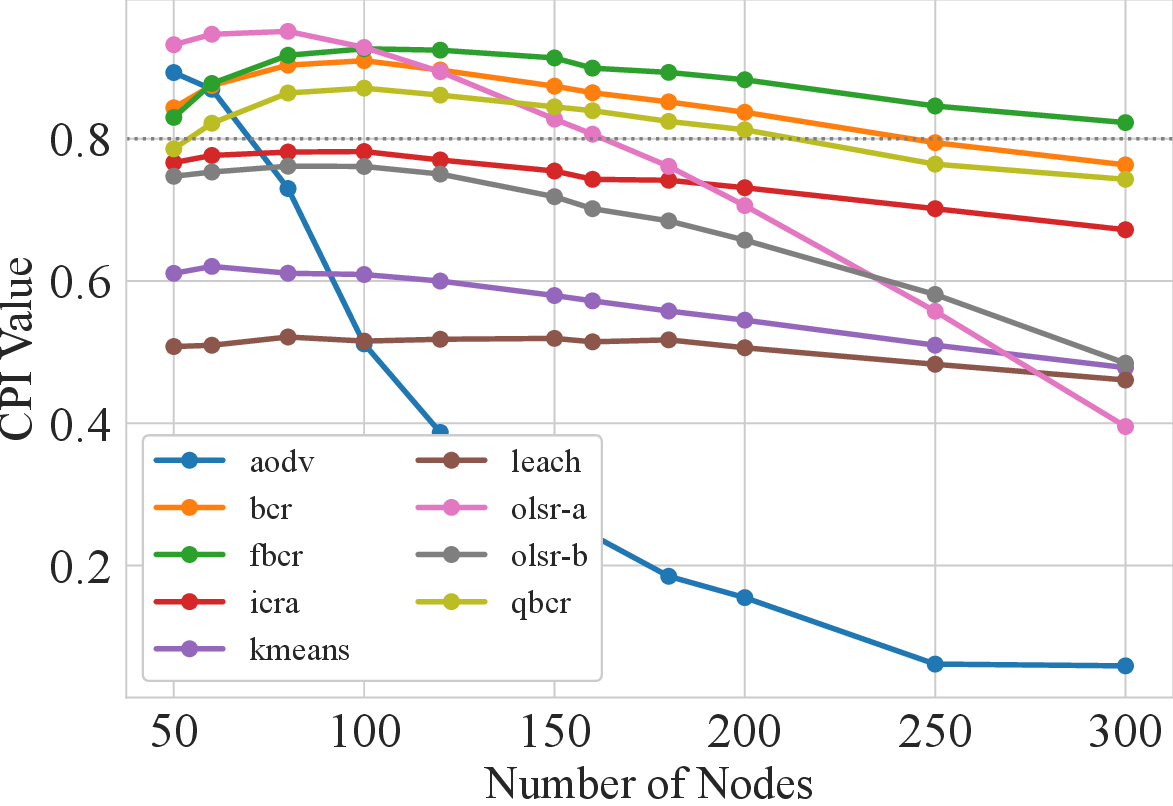}%
\label{fig:cpi_scale}}
\hfil
\subfloat[]{\includegraphics[width=2in]{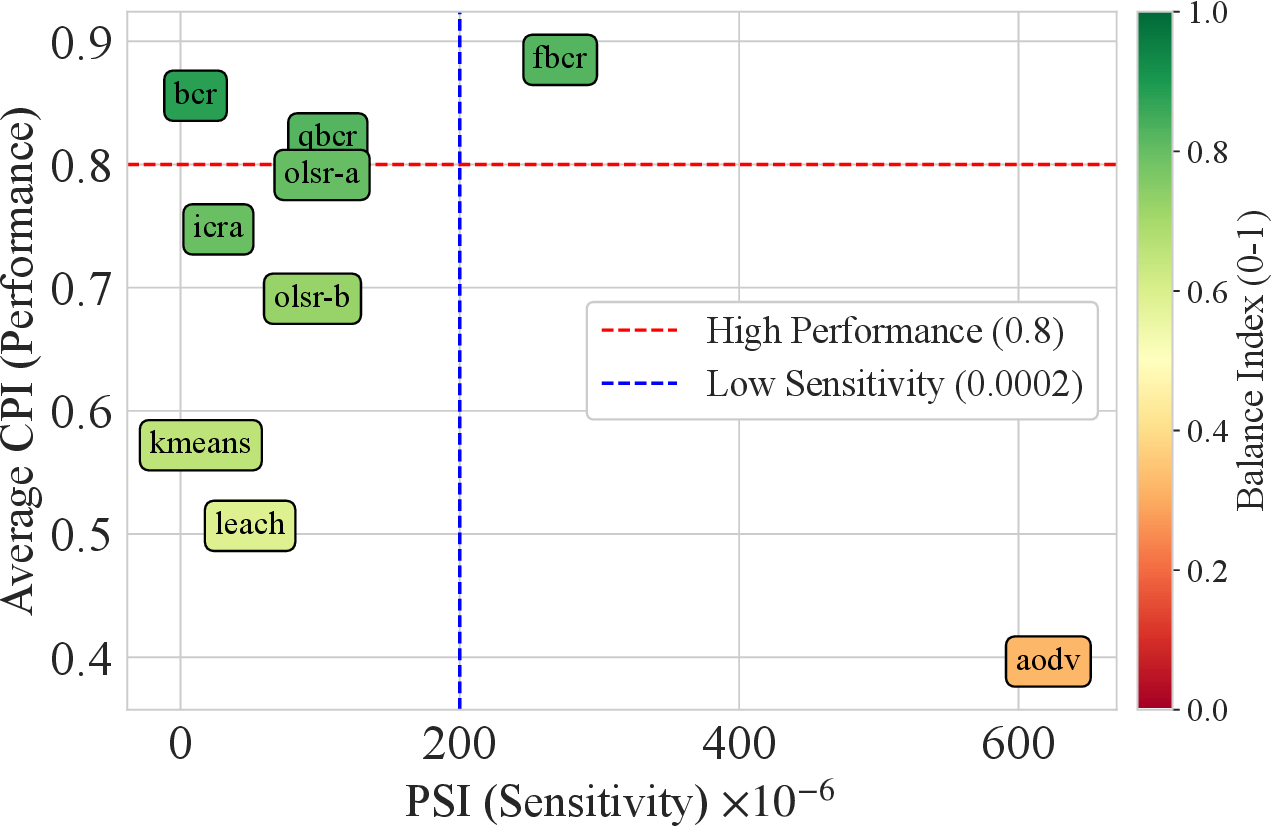}%
\label{fig:psi_scale}}
\hfil
\subfloat[]{\includegraphics[width=2in]{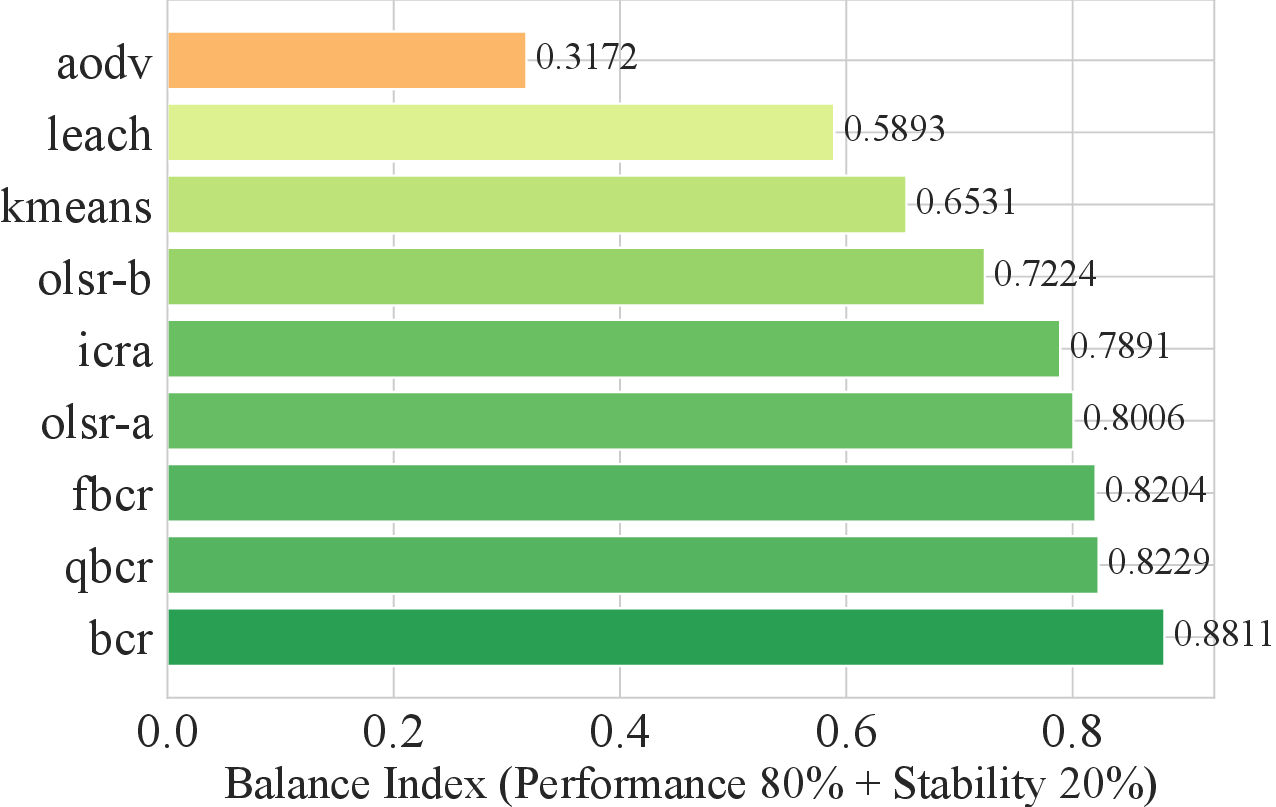}%
\label{fig:balance_scale}}

\subfloat[]{\includegraphics[width=2in]{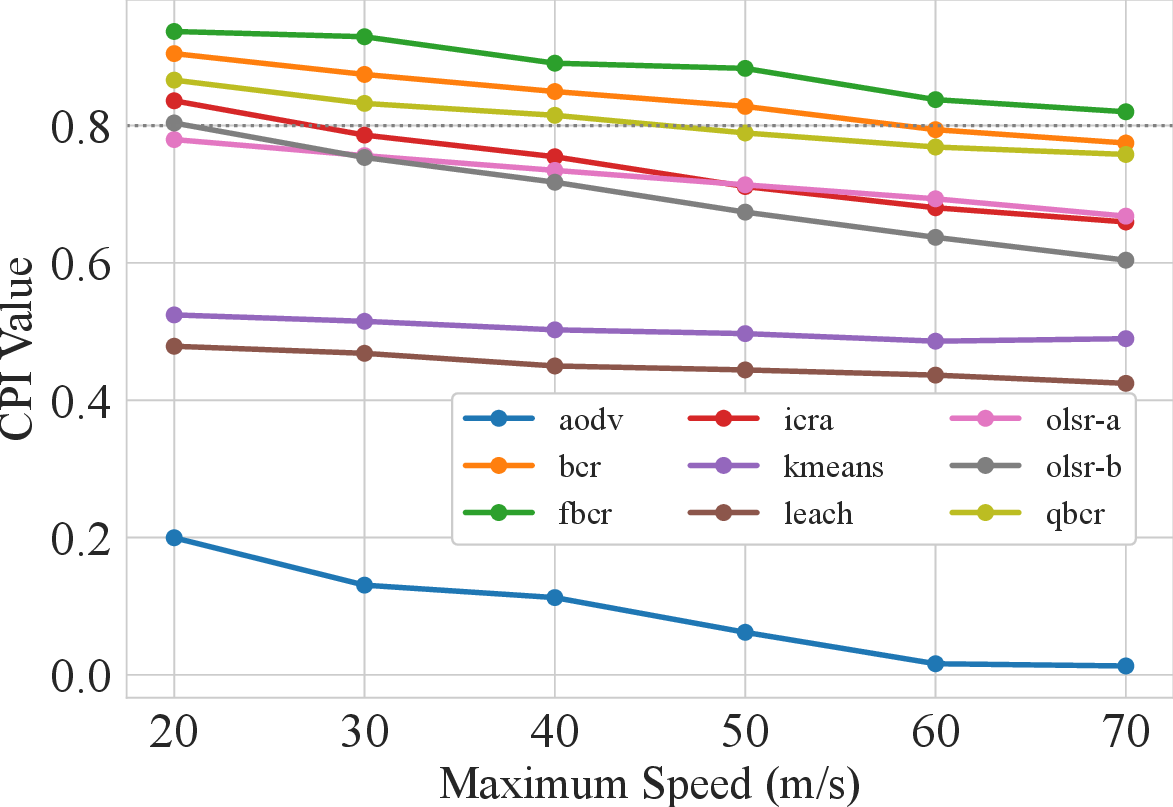}%
\label{fig:cpi_speed}}
\hfil
\subfloat[]{\includegraphics[width=2in]{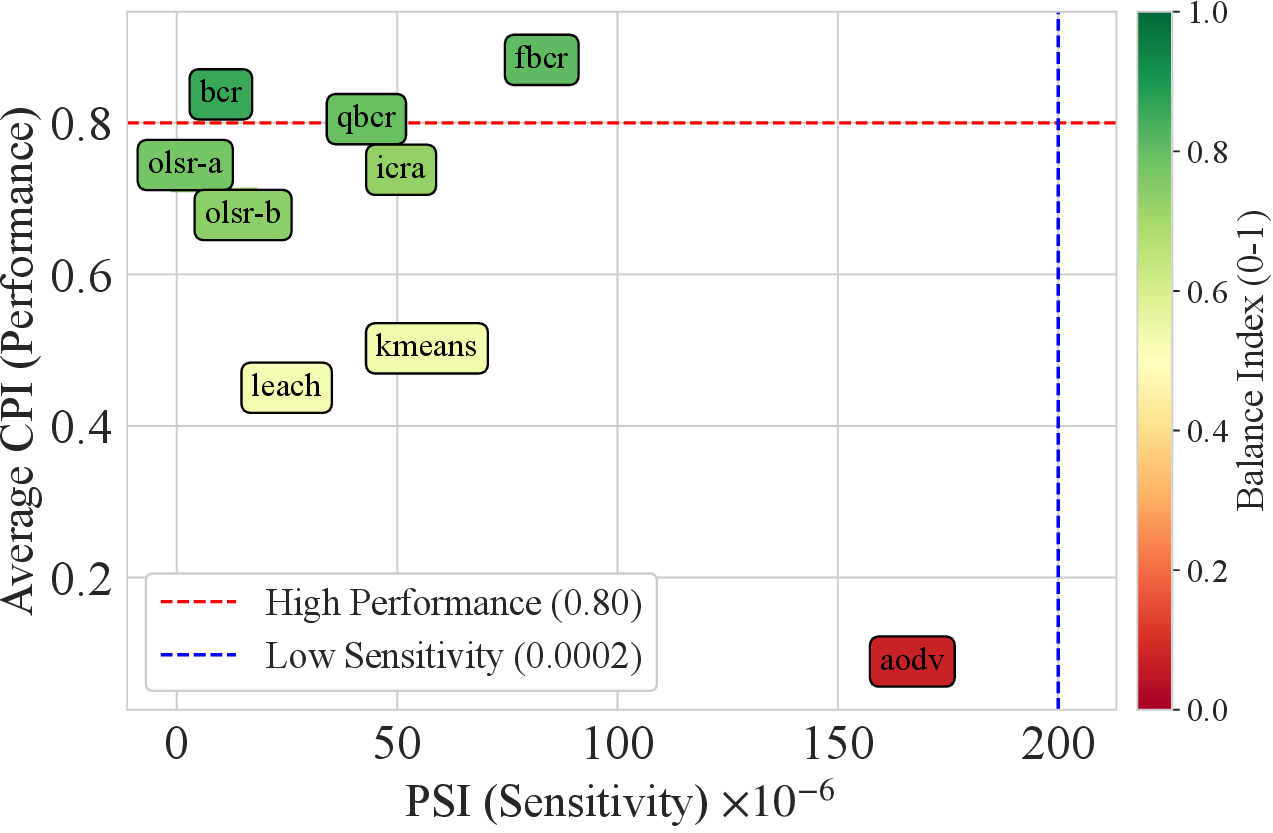}%
\label{fig:psi_speed}}
\hfil
\subfloat[]{\includegraphics[width=2in]{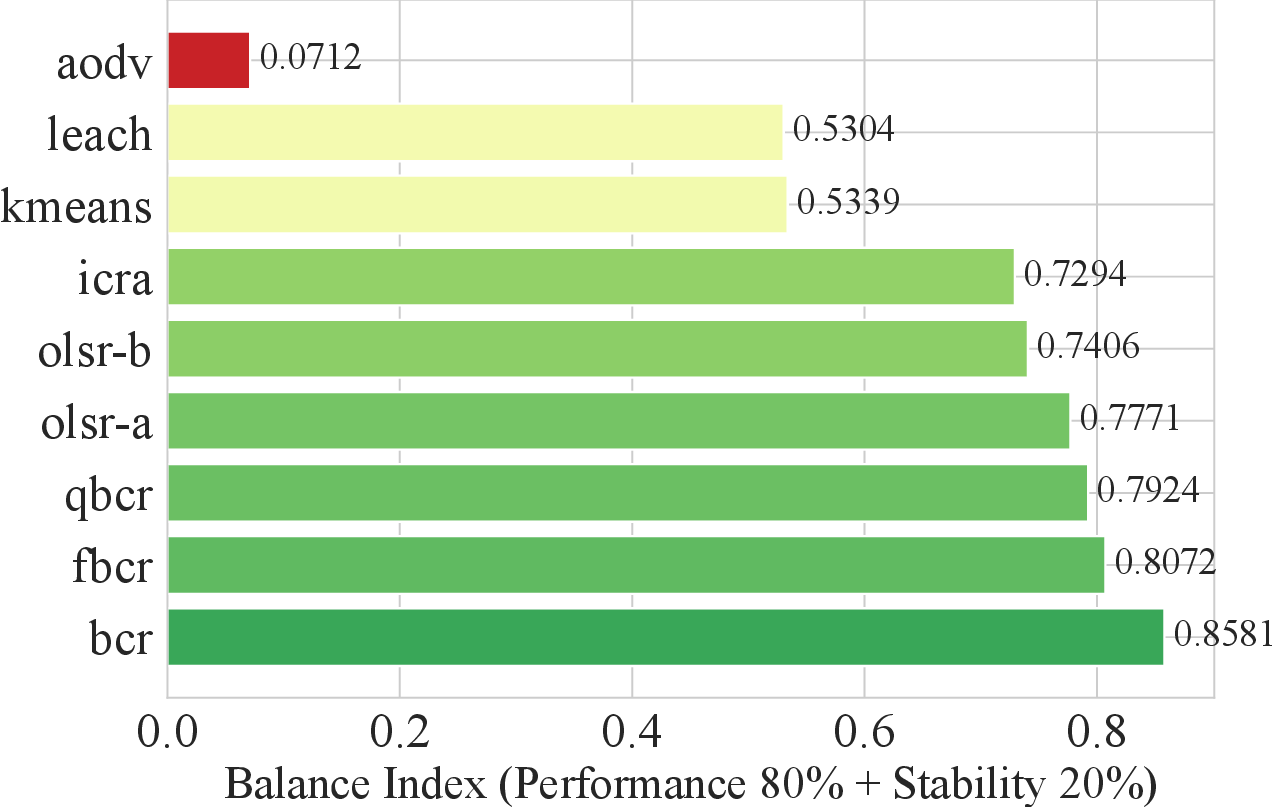}%
\label{fig:balance_speed}}
\caption{Comparison of PSI and CPI based on different node scale and speed. (a) CPI trends vary by node scale. (b) Performance–resilience (CPI–PSI) distribution under node scale variation. (c) Comprehensive Ranking of Node Scale variant. (d) CPI trends vary by node speed. (e) Performance–resilience (CPI–PSI) distribution under node speed variation. (f) Comprehensive Ranking of Node Speed variant.}
\label{fig:combined_in_one}
\end{figure*}

Fig.~\ref{fig:combined_in_one} presents the sensitivity analysis results of different routing schemes under variations in network node scale (\subref{fig:cpi_scale}, \subref{fig:psi_scale} and \subref{fig:balance_scale}) and node mobility range (\subref{fig:cpi_speed}, \subref{fig:psi_speed}, \subref{fig:balance_speed}). From the CPI trend curves in Fig.~\ref{fig:combined_in_one}\subref{fig:cpi_scale} and \subref{fig:cpi_speed}, the proposed xBCR family consistently outperforms the baseline protocols under both scenarios. When the number of nodes exceeds 100, FBCR achieves the highest CPI, while BCR and QBCR stably rank second and third. Traditional non-clustered protocols such as AODV and OLSR experience rapid CPI degradation as the network scales, indicating limited scalability. Cluster-based baselines (LEACH, K-means, ICRA) are relatively stable but with lower overall CPI.

%-------------------PSI_CPI_Balance---------------------

Fig.~\ref{fig:combined_in_one}\subref{fig:psi_scale} and \subref{fig:psi_speed} visualize each protocol as a point in the performance--resilience plane, with the horizontal axis denoting sensitivity (PSI, scaled by $10^{6}$) and the vertical axis the mean composite performance $\overline{CPI}$. A protocol thus sits higher when its overall performance is stronger and farther left when it is less sensitive to parameter variation; the dashed reference lines ($\overline{CPI}=0.80$ and the low-sensitivity threshold) partition the plane into four quadrants, of which the upper-left is most desirable. Marker colors encode the balance index $\beta$ in \eqref{eq:bi} on a red-to-green scale. BCR resides in the upper-left region, achieving both strong performance and stable behavior, whereas AODV falls in the lower-right. Notably, FBCR attains the highest $\overline{CPI}$ but shifts rightward (larger PSI); we attribute this to the heterogeneous hello update frequencies introduced by the fuzzy logic optimizer, which exposes a trade-off between peak performance and parametric robustness.
%show the CPI--PSI distribution, where greater distance from the X-axis indicates higher performance and closer proximity to the Y-axis implies better stability. BCR achieves both strong performance and stable behavior, whereas AODV performs poorly on both dimensions. Notably, FBCR exhibits the highest overall performance but shows increased sensitivity, which we attribute to heterogeneous hello update frequencies introduced by the fuzzy logic optimizer. This suggests a trade-off between peak performance and robustness in practical deployments.

Fig.~\ref{fig:combined_in_one}\subref{fig:balance_scale} and \subref{fig:balance_speed} rank the protocols by the balance index of \eqref{eq:bi}. The choice of $w_p = 0.8$ and $w_s = 0.2$ follows the Pareto principle (the ``80/20 rule''): the average performance $\overline{CPI}$ is taken as the dominant contributor, while stability $\mu$ is retained as a $20\%$ penalty term for protocols with unstable behavior. Table~\ref{tab:weight-sweep} reports protocol ranks under different $w_p$. BCR/QBCR and the AODV/LEACH/K-means remain in the top and bottom tiers across all configurations, respectively. Although FBCR could achieve the highest absolute CPI in most scenarios, the rank of $\beta_{fbcr}$ shifts with $w_p$, highlighting the inherent trade-off between absolute performance and resilience.

\begin{table}[!t]
\centering
\caption{Protocol ranks under $\beta_q$ for varying $w_p$ ($w_s = 1 - w_p$)}
\label{tab:weight-sweep}
\setlength{\tabcolsep}{3.5pt}
\renewcommand{\arraystretch}{1.15}
\begin{tabular}{lccccccccccc}
\toprule
& \multicolumn{5}{c}{Node-scale} & & \multicolumn{5}{c}{Speed-variation} \\
\cmidrule(lr){2-6} \cmidrule(lr){8-12}
$w_p$ & 0.5 & 0.6 & 0.7 & 0.8 & 0.9 & & 0.5 & 0.6 & 0.7 & 0.8 & 0.9 \\
\midrule
BCR     & \textbf{1} & \textbf{1} & \textbf{1} & \textbf{1} & \textbf{1} & & \textbf{1} & \textbf{1} & \textbf{1} & \textbf{1} & \textbf{1} \\
FBCR    & \textit{7} & 5          & 5          & \textbf{3} & \textbf{2} & & 6          & 5          & 4          & \textbf{2} & \textbf{2} \\
QBCR    & \textbf{3} & \textbf{3} & \textbf{2} & \textbf{2} & \textbf{3} & & 4          & 4          & \textbf{3} & \textbf{3} & \textbf{3} \\
ICRA    & \textbf{2} & \textbf{2} & \textbf{3} & 5          & 5          & & 5          & 6          & 6          & 6          & 5          \\
OLSR-a  & 4          & 4          & 4          & 4          & 4          & & \textbf{2} & \textbf{2} & \textbf{2} & 4          & 4          \\
OLSR-b  & 6          & 6          & 6          & 6          & 6          & & \textbf{3} & \textbf{3} & 5          & 5          & 6          \\
K-means & 5          & \textit{7} & \textit{7} & \textit{7} & \textit{7} & & \textit{8} & \textit{8} & \textit{8} & \textit{7} & \textit{7} \\
LEACH   & \textit{8} & \textit{8} & \textit{8} & \textit{8} & \textit{8} & & \textit{7} & \textit{7} & \textit{7} & \textit{8} & \textit{8} \\
AODV    & \textit{9} & \textit{9} & \textit{9} & \textit{9} & \textit{9} & & \textit{9} & \textit{9} & \textit{9} & \textit{9} & \textit{9} \\
\bottomrule
\end{tabular}
\end{table}

\section{Conclusion and Future Works}\label{sec:conclusion}

This paper proposed FBCR, a robust routing scheme designed for highly dynamic FANETs. To ensure structural stability and adaptability, we introduced a dual-layer beacon mechanism that enables efficient proactive cluster maintenance, coupled with a fuzzy logic-based hello optimizer that dynamically regulates control overhead to match network volatility. Furthermore, we formulated the resilience metric PSI to quantitatively evaluate protocol robustness against environmental perturbations. Extensive simulations demonstrate that in large scale and dynamic networks, FBCR and BCR outperform benchmark protocols in both communication performance and parametric resilience.

Future work will focus on developing more customized evaluation metrics, such as network healing capability, to better assess protocol reliability under failure conditions. Additionally, we plan to extend the framework to multi-layer heterogeneous UAV networks and explore deep reinforcement learning for intelligent cross-layer optimization.

\section*{Acknowledgment}
This work is supported by the National Natural Science Foundation of China (Grant No. 62372462).

\bibliographystyle{IEEEtran}
\bibliography{references}

\end{document}